# GESSURE: A ROBUST FACE-AUTHENTIC ENABLED DYNAMIC GESTURE RECOGNITION GUI APPLICATION

Ankit Jha, Ishita, Pratham G. Shenwai, Ayush Batra,
Siddharth Kotian and Piyush Modi

Manipal Institute of Technology, MAHE, Manipal, India

## ABSTRACT

*Using physical interactive devices like mouse and keyboards hinders naturalistic human-machine interaction and increases the probability of surface contact during a pandemic. Existing gesture-recognition systems do not possess user authentication, making them unreliable. Static gestures in current gesture-recognition technology introduce long adaptation periods and reduce user compatibility. Our technology places a strong emphasis on user recognition and safety. We use meaningful and relevant gestures for task operation, resulting in a better user experience. This paper aims to design a robust, face-verification-enabled gesture recognition system that utilizes a graphical user interface and primarily focuses on security through user recognition and authorization. The face model uses MTCNN and FaceNet to verify the user, and our LSTM-CNN architecture for gesture recognition, achieving an accuracy of 95% with five classes of gestures. The prototype developed through our research has successfully executed context-dependent tasks like save, print, control video-player operations and exit, and context-free operating system tasks like sleep, shut-down, and unlock intuitively. Our application and dataset are available as open source.*

## 1. INTRODUCTION

The post-pandemic world adopted a contactless approach to reduce the risk of infection. Human-Computer Interaction technologies that eliminate the need to touch any surface provides innovative answers to new issues as COVID-19 continues to transform our daily lives, resulting in a safer society. Hand gestures are the most notable of all body motions, which include the actions of the arms, fingers, or hands. It is possible to use a comprehensive communication system with various possible movements. The advancement of ubiquitous computing in the modern era has made using natural user interfaces essential, eliminating the use of devices like mouse and keyboards. Early gesture recognition applications used glove-based systems [1] and shifted towards visual systems using depth cameras [32]. Current research has eliminated the need for external hardware and uses sophisticated deep learning architecture that can utilize the raw feed from the webcam in our laptops and computers.

There are two broad categories of hand gestures: static and dynamic. A static gesture is an arrangement of the hands in a specific stance represented by a single image. In contrast, a dynamic hand gesture consists of a hand movement trajectory, and the hand shape changes between a start or trigger time point to an endpoint. It is less challenging to develop systems that recognize static gestures than dynamic ones, as segmenting a static frame is more accessible than a dynamic frame. However, from the user experience viewpoint, static gestures are less intuitive and require the user to remember the standard gestures to operate the gesture-based application tasks. The use of static hand gestures limits hand motion. Dynamic hand movements make it easier for users to communicate. As part of this research, the aim is to develop a system that incorporates intuitive, dynamic gestures and even allows the user to define their own pair of gestures that can be trained on the model and used as macros when performing tasks.

In the field of HCI, user authentication is becoming increasingly important. It is a process that enables a device to verify a person's identity when interacting with the interface and only permits the designated user to control the system's functionalities. The development of gesture recognition systems for manipulating computer functions and frequently used tasks such as printing, saving, screenshots, and sleeping has been extensive in the past. They use computer vision techniques utilizing the video feed from the laptop or PC webcam, lacking a biometric security feature that will allow the pre-verified user to utilize the system's gesture manipulation abilities. Our paper proposes a face verification layer before executing the gesture recognition task. Our prototype integrates the face verification module and the gesture recognition module into a GUI-based application which functions as a stand-alone application for gesture-based manipulation tasks. Currently, the system recognizes five gestures and executes macros based on them. It can execute the following tasks using dynamic gestures: print, save, exit (in-application tasks in MS-Word), screen lock, screen-unlock, system restart, and shut-down (Operating System tasks). Our prototype prioritizes extensibility, allowing us to assign specific gestures to different tasks using macros. It also allows the user to train their own set of gestures.

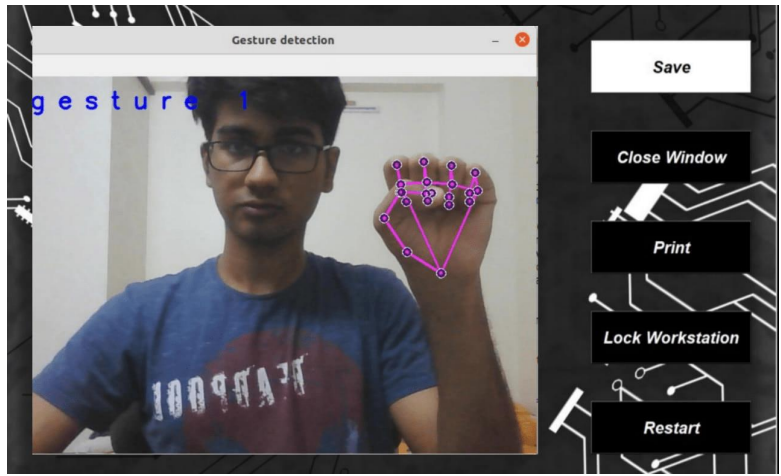

Fig.1. GesSure UI Mock-up

## 2. RELATED WORKS

### 2.1. Face Verification

For biometric security applications, face verification usually works as two successive systems, the first being a face recognition system, localizing the human face, and a verification system that verifies whether two images belong to the same subject. Face verification has been a broad subject of study in computer vision, with ongoing research and development improved algorithms to boost robustness and reliability. Geometry-based and appearance-based approaches are common in traditional approaches. Geometry-based features describe the contour of the face and its components, such as the mouth or the eyebrow, while appearance-based features explain the texture of the face generated by the expression [14]. Feature-based methods using SURF [4] and SIFT [31] are scale-invariant, and the method works in two steps- detection of the feature point and feature description. In order to categorize areas of an image, techniques like Haar-Cascade take consecutive rectangular regions in a detection window at a specified point in the picture, add up the pixel intensities in each sector, and then compute the difference between these sums, introduced by Viola and Jones [38]. Use of hybrid methods involving feature extraction using SIFT and, after that, projecting them into a subspace with a lower dimension using PCA [23]. Modern systems use neural networks with CNNs with extensively trainable parameters for computer vision tasks. CNNs use sophisticated gradient descent and a variety of choices for loss function as an optimizer to the model [12]. Usually, CNNs perform poorly under a constrained training dataset condition; thus, in this paper, we have approached the problem by combining the best of deep learned features with a traditional One-Shot learning framework[39][8]. We designed our face detection model using MTCNN [42], feature extraction using FaceNet [34], and Softmax classification.

## 2.2. Gesture Detection

In this unit, we discuss previous work related to gesture detection. Major works related to this field started with the advancement in HCI. Different methods using wearable gloves were proposed [7]. It used various sensors like the curvature sensor [24], accelerometer sensor [15], optical fibre transducer [13], flex sensor [36], and angular displacement sensor [16]. However, these methods require expensive hardware setup and prove to be cumbersome to handle. It also restricts natural hand movement, which is especially inconvenient for those with chronic illness or the elderly. The vision-based approach tends to overcome these particular challenges. A study by [25], [40] explored the computer-vision approach using marked, colored gloves. It was still necessary to wear gloves for this.

More modern approaches involve employing Computer Vision without any gloves. A variety of cameras can be utilized, such as TOF [5, 28] , monocular [43], Kinect [41, 21], RGB [21] and Depth Camera [41, 10, 21]. however, this has various challenges like lighting variation, the effect of occlusions, complex background, resolution against processing time, or color matching with the background. Computer vision has seen many variations, such as color-based recognition (pixel-based skin detection is not very reliable [6]). Appearance-based analyses extract features to visualize objects like a hand; they include the location of palm, location of fingertips, orientation, and direction [20, 30], skin color [37], and eigenvectors [35], among many more. Models like [19] studied the possibility of 3D models; however, these were highly complex and unsuitable for general use. Like [28], motion-based recognition utilized movement modeling and pattern recognition. However, it faced issues when multiple gestures were active during the recognition period. Skeleton-based recognition systems [21] focus on geometric and statistical features like joint orientation, the space between joints, the skeletal joint location, and more. Here, the range of detection proves to be the major challenge. Other techniques like separating background and foreground [33] and background subtraction have proved less accurate than desired. The deep-learning-based approach appears reliable because of the learning role principle. It uses multilayers of data for learning and giving accurate predictions. Hidden Markov Models (HMMs) [20], K Nearest Neighbor (KNN), and Support Vector Machines (SVMs) [9] are among the tested models. The use of neural networks [37], LSTM [27], RNN [22], and 3D-CNN combined with LSTM [29], Adapted Deep Convolutional Neural Network (ADCNN) [2], deep CNN [3], LRCN [18] have also shown much higher success rates than their predecessors. Most of these methods require large datasets, which limits the gestures and makes them difficult to remember for the user.

In our approach, for hand detection, we use an open-source library, namely MediaPipe. It can effectively extract the coordinates of landmarks on the hand irrespective of the lighting conditions, palm size, orientation, and distance from the camera. A depth sensor or 3D camera is no longer needed. For gesture recognition, we use CNN and LSTM to extract spatial and temporal features, allowing us to use past information to train the following layers. LSTM is used to make time-series data predictions, which helps collect temporal features, while CNN collects spatial features and makes predictions about the landmarks. We allow the user flexibility to add gestures as per his wish, thus making it more intuitive. Another highlight of our approach is the increased safety feature. We use a Deep Learning based face recognition model that verifies the user and responds only to their gesture actions. It uses one-shot learning, thus giving a high accuracy with even a tiny dataset irrespective of background conditions.

## 3. METHODOLOGY

### 3.1. Dataset

#### 3.1.1. Face Verification

For face verification, we have used OpenCV to make our dataset. With a Wait-Key of 500 milliseconds, each person can record for 20 frames. It is necessary to incorporate as many angles as possible to achieve the best possible accuracy. We also added an extra category, namely "stray user," where random faces are stored so that the model can differentiate between the user and the stray. For our assignment, we have one user, and the stray user category has four entries.

#### 3.1.2. Gesture Detection

The dataset consists of 5 gestures pulled from 4 people and stored in the form of NumPy arrays. Random gestures have been added as an extra category so that the model can recognize random gestures more effectively. Each person has recorded each gesture for 20 frames, and all the data stocked together has been used to train our model. We have used NumPy, Matplotlib, OpenCV, and the library MediaPipe to achieve this.

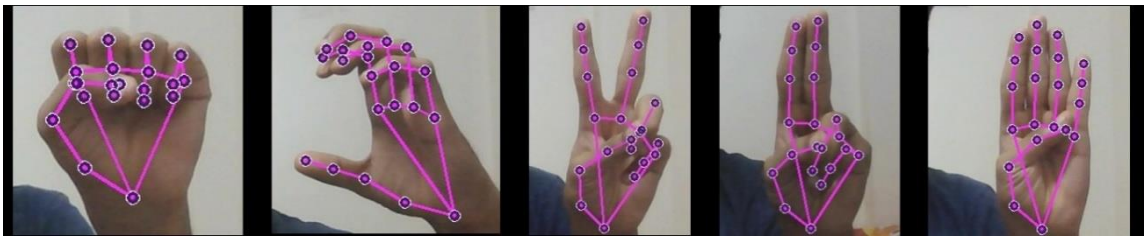

Fig.2. Hand Gesture examples

### 3.2. Architecture

#### 3.2.1. Face Verification

We will concentrate on well-liked deep learning methods, performing Softmax classification, FaceNet [34] feature extraction, and MTCNN [42] face identification [11] [17]. A neural network called MTCNN, or Multi-Task Cascaded Convolutional Neural Networks, is used to recognize faces and other facial landmarks in photographs. FaceNet is a deep neural network that identifies features in pictures of faces. We have used MTCNN to detect faces in the input image and then passed the detected faces to FaceNet to generate and categorize face embeddings. FaceNet receives a picture as input and gives a face embedding vector of 128 numbers. This embedding contains essential features of the face in an image. We can plot an image in the cartesian coordinate system using the embeddings generated from FaceNet. Models usually attempt to generate face embeddings for unseen face images and then compare them with trained data embeddings. The final allocation goes to the class with the closest embedding representation to the testing image.FaceNet learns the information about the face in the following way:

(1) Randomly select an anchor image.
(2) Randomly select an image of the same person as the anchor image (positive example).
(3) Randomly select an image of a person different than the anchor image (negative example).
(4) Adjust the FaceNet network parameters so that the positive example is closer to the anchor than the negative one.

*Triplet loss* is a learning strategy that uses an anchor, positive and negative instances. The output embeddings of the FaceNet are then passed on to classifiers such as KNN, Softmax, Random Forest, and SVM, depending on the use case. Here, we have used Softmax because it can work well for the neural network.

Triplet Loss: serves as the embedding's representation. In a d-dimensional Euclidean space, it embeds an image x. We further restrict this embedding to exist only on the d-dimensional hypersphere. Using the nearest-neighbor method, we calculate the loss. Here we want to ensure that an image (anchor) of a specific person is closer to all other images (positive) of the same person than it is to any image (negative) of any other person. Triplets that are simple to satisfy would result from creating all feasible triplets. The network would still route these triplets, preventing them from being used in training and hindering convergence. Choosing tricky triplets that are engaged and can advance the model is critical.

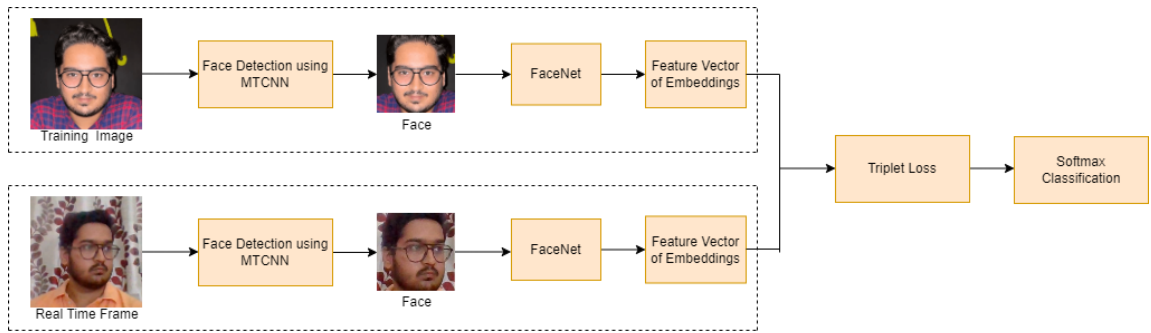

Fig.3. Architecture of Face Verification.

### 3.2.2. Gesture Detection

### 3.2.2.1. Detection and Localization of Hands

MediaPipe [26] is a framework that enables developers to build multi-modal cross-platform applied ML pipelines. They track key points on different body parts as the skeleton of nodes and edges or landmarks. All coordinate points are three-dimension normalized. The Hand tracking system comprises a backend machine learning pipeline of two interdependent models: Models for palm and hand landmarks detection. In addition, the palm detection model provides the landmark model with an accurately cropped palm image. This process diminishes data augmentation (i.e., Rotations, Flipping, Scaling) done in Deep Learning models and dedicates most of its power to landmark localization. In this Palm Detector, using the ML pipeline challenges a different strategy.

Detecting hands is a complex procedure as we have to perform image processing and thresholding and work with various hand sizes, leading to time consumption. Instead of directly detecting hands from the current frame, first, the Palm detector is trained, which estimates bounding boxes around rigid objects like palms and fists, which is more straightforward than detecting hands with coupled fingers. Secondly, an encoder-decoder extracts a larger scene context. Following the detection of palms across the entire image frame, subsequent Hand Landmark models enter the picture. This model precisely localize 21 3D hand-knuckle coordinates (i.e., x, y, z-axis) inside the detected hand regions. The model is so well trained and robust in hand detection that it maps coordinates to partially visible hands. Now that we have a working Palm and Hand detection model, we ask the user to demonstrate the hand gestures in front of the webcam.

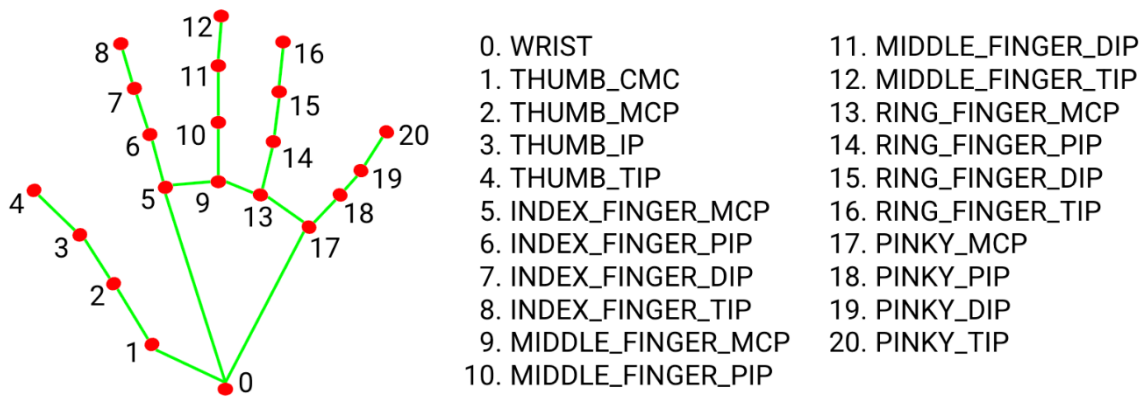

Fig.4. Hand Landmarks in MediaPipe [26]

**3.2.2.2. Identification of Hand Gestures**

The proposed deep Conv-LSTM neural network comprises two convolutional layers, one LSTM layer, four dropout layers, one time-distributed layer, and one dense layer. The convolutional and LSTM layers use Rectified Linear Unit as the activation layer, while the final dense layer uses the function as activation. The entire architecture consists of 237,718 parameters. The architecture begins with an initial convolutional layer of size 3x3. Considering the gesture is made up of 20 frames, with each frame containing x, y, and z coordinates of the 21 landmarks, we take the input size to be 20x63. A max-pooling layer of pool size two extracts the most prominent features from the output. We then use a dropout layer to reduce the overfitting of data. There is a 25% dropout size set. Then the output goes through a similar convolutional layer followed by a max-pooling and two dropout layers. The output from these above layers is passed through a time-distributed layer, allowing us to layer every temporal slice of an output. This layer allows us to pass the output from the convolutional layers into the LSTM layer. The LSTM layer is composed of 200 neurons with the rectified linear units as the activation function. We used only a single LSTM layer as we achieved significant accuracy with only one layer avoiding the need to utilize additional LSTM layers. A dropout layer of 0.25 size follows the LSTM layer to further reduced overfitting from the LSTM layer. Finally, we pass the output through a dense layer with Softmax activation.

International Journal on Cybernetics & Informatics (IJCI) Vol. 11, No.4, August 2022

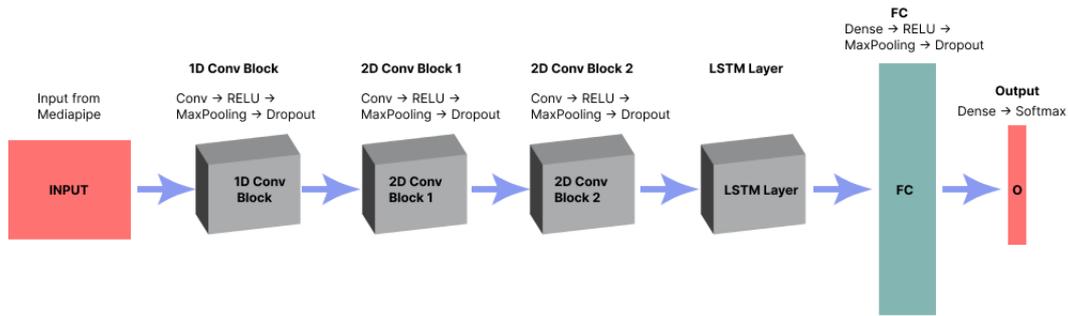

Fig.5. Architecture of Gesture Detection

## 3.3. GUI Prototype Approach

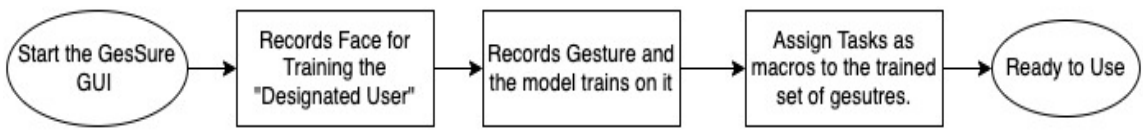

Fig.6. Flowchart of starting GesSure in the training phase

GesSure's prototype uses a training and in-use approach. In the training phase, the GUI provides an interface for the user to train the Face Verification model to assign the "designated user," which is a mandatory step when registering for the first time. It records the face of the user; the model takes 20 images, preferably from different angles, stores them in the user directory, and sets the path as the verified user. The GUI then provides an interface allowing users to record or use the pre-trained gestures. One can also assign existing tasks as macros to the gestures. The gestures are dynamic and provide a better user experience with the possibility of intuitive gestures that relieves the user from remembering standard gestures. The camera captures 20 frames for each gesture, from start to end motion. The macro-options currently in the prototype are the frequently used tasks - save, exit, print, screen-lock, screen unlock, system shut down, and system restart. Save, print, and exit operations are context-sensitive, which applies to current applications. Using the Save gesture on a word document, for example, saves the document; using the Print gesture opens the printer dialog, and so on. Likewise, allocating a gesture for the command close/exit is also possible, which terminates the current application. The system will shut down if no applications are open. It is similar to Alt+F4 key press functionality on windows PC.

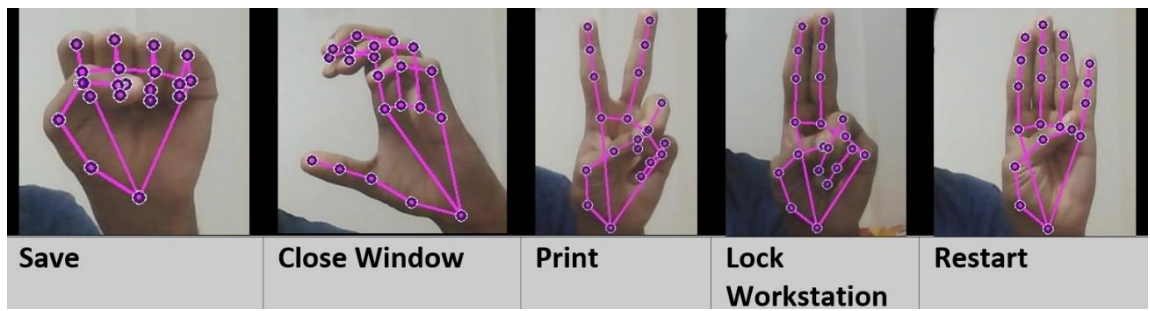

Fig.7. Hand Gestures and their designated tasks in MS-Word.

In the In-Use phase, the user can activate the GesSure prototype with a toggle button. The prototype uses the face verification model trained in the training phase as a security overlay with the user's face images. The gesture recognition module can be activated only when the face verification algorithm verifies the user's presence. The algorithm [1] uses One-shot learning, with MTCNN[2] detecting the face and FaceNet[3] extracting facial features and a softmax layer for classification. The verified user can now manipulate the tasks assigned to the trained gesture motions. Here, we use CNN combined with LSTM. We use three convolution layers (1-dimensional and 2-dimensional) to extract spatial features. In order to extract temporal features, we use LSTM layers since they are good at storing knowledge from previous and current frames. It is more suitable than GRU because of its lightweight, and our use case required a limited time sequence. Dropout layers prevent the model from overfitting and make it suitable for generalized use.

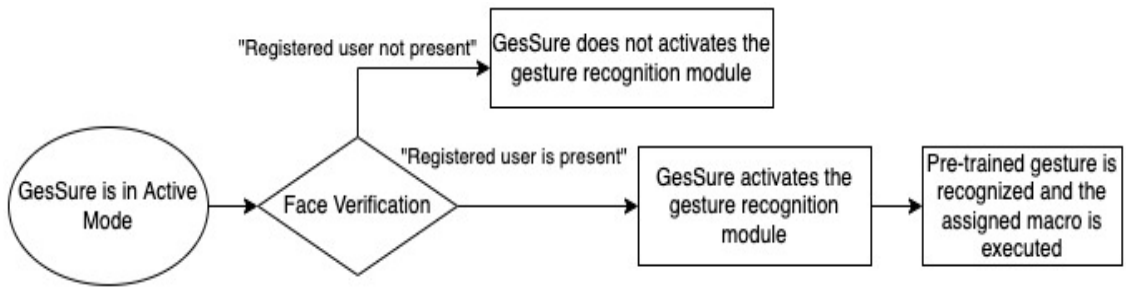

Fig.8. Flowchart of GesSure in In-Use Phase.

## 4. RESULTS

### 4.1. Face Verification

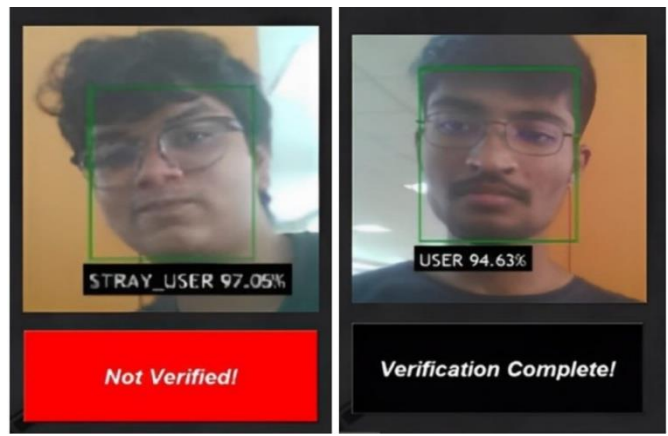

Fig.9. Mock-up UI of Face Verification

The face verification model achieves high accuracy and classifies a non-registered user as a "stray user." The application activates the gesture recognition module only when the user detection accuracy is above 90%. The model also accepts the user if there are multiple people in the frame and the user is among them.

**4.2. Gesture Detection**

```
              precision    recall  f1-score   support

           0      0.923     1.000     0.960        12
           1      1.000     1.000     1.000        12
           2      1.000     0.857     0.923        14
           3      0.909     1.000     0.952        10
           4      1.000     0.833     0.909         6
           5      0.857     1.000     0.923         6

    accuracy                          0.950        60
   macro avg      0.948     0.948     0.945        60
weighted avg      0.955     0.950     0.949        60
```

Fig. 10. Accuracy of Gesture Detection

The gesture recognition model uses MediaPipe for localizing hand coordinates, and our LSTM-CNN model achieves an accuracy of 95% on the five gesture classes.

**4.3. Task Manipulation**

We trained the five gestures and assigned several tasks as a macro to those gestures. There are two broad categories of tasks: context-dependent and context-free. Gestures in context-dependent tasks can trigger application-level tasks. In MS Word, we added gesture trigger tasks like save, print and exit. In VLC Media Player, we added tasks like seek, play/pause, and volume controls. Tasks triggered by gestures are context-free and associated with operating systems, such as restarts, screen lock, and shut-downs.

**4.4. Code and Dataset Availability**

Code for the models, the GUI and the data repositories are made available as Open-Source on GitHub (https://github.com/PrathamShenwai/GesSure/tree/master).

**5. CONCLUSION AND FUTURE SCOPE**

The hand gesture recognition system has significantly impacted the development of effective human-machine interaction. Such systems are an accessibility aid to differently abled persons who can utilize hand movements to communicate. The current state of gesture recognition systems has certain shortcomings. They are unsafe as they do not integrate any authentication check. Also, they are less intuitive since they use static gesture techniques like the American sign language models. This paper focused on building a prototype of a user authentication integrated HCI system that assists users in manipulating frequently used context-free and operating system tasks using a gesture recognition system. Instead of going for "difficult to remember and easy for the algorithm to classify," we use meaningful and relevant gestures for task operation, resulting in a better UX. For future work, we plan to work on the extensibility of the prototype, enhance the user interface and develop it as an open-source package that can be downloaded and used on the go on several operating systems.

Additionally, we can expand the system to be beyond personal computers. These gesture assistance tools can be integrated into ATMs, improving the user experience for differently abled users. Another use case is automated dashboards in vehicles, which can enable the user to interact with the console meanwhile minimizing the necessity of contact.